\documentclass[letterpaper, 10 pt, conference]{ieeeconf}  
\IEEEoverridecommandlockouts                              
\usepackage{amssymb, bm}
\usepackage{textcomp}
\usepackage[T1]{fontenc}        
\usepackage[utf8]{inputenc}		
\usepackage{fancyhdr}
\usepackage{microtype}

\usepackage{booktabs,multirow,siunitx}

\usepackage{cite}
\usepackage{url}

\usepackage{graphicx}
\usepackage{xcolor}
\usepackage[caption=false,subrefformat=parens,labelformat=parens]{subfig}
\usepackage{comment}

\usepackage{amsmath,amssymb}
\usepackage{breqn}

\usepackage{algorithm}
\usepackage{algpseudocode}
\usepackage[authormarkup=superscript,deletedmarkup=sout,addedmarkup=em]{changes}
\usepackage{soul}
\definechangesauthor[name={Constantinos  Antoniou}, color=green!75!black]{CA}
\definechangesauthor[name={Rafał Kucharski}, color=orange!75!red]{RK}
\definechangesauthor[name={Guido Cantelmo}, color=orange!60!red]{GC}
\setremarkmarkup{\footnote{\textcolor{Changes@Color#1}{#1:~#2}}}
\soulregister\cite7
\soulregister\ref7
\soulregister\pageref7

\newcommand{\mytitle}[1]{\smallskip\noindent\textbf{#1:}}

\newcommand{\eq}[1]{(eq.\ref{#1})}

\usepackage[caption=false,subrefformat=parens,labelformat=parens]{subfig}
\title{\LARGE \bf
A low dimensional model for bike sharing demand forecasting
}

\author{ \parbox{2 in}{\centering Cantelmo Guido \\
         Department of Civil, Geo and Environmental Engineering\\
         Technical University of Munich\\
         Munich, Germany\\
         {\tt\small g.cantelmo@tum.de}}
         \hspace*{ 0.2 in}
         \parbox{2 in}{ \centering Kucharski Rafał\\
         Department of Transportation Systems\\
         Cracow University of Technology\\
         Cracow, Poland\\
         {\tt\small rkucharski@pk.edu.pl}}
         \hspace*{ 0.2 in}
         \parbox{2 in}{\centering Antoniou Constantinos\\
         Department of Civil, Geo and Environmental Engineering\\
         Technical University of Munich\\
         Munich, Germany\\
         {\tt\small c.antoniou@tum.de}}
}

\begin{document}

\maketitle
\thispagestyle{empty}
\pagestyle{empty}

\begin{abstract}
Big, transport-related datasets are nowadays publicly available, which makes data-driven mobility analysis possible. Trips with their origins, destinations and travel times are collected in publicly available big databases, which allows for a deeper and richer understanding of mobility patterns.

This paper proposes a low dimensional approach to combine these data sources with weather data in order to forecast the daily demand for Bike Sharing Systems (BSS). The core of this approach lies in the proposed clustering technique, which reduces the dimension of the problem and, differently from other machine learning techniques, requires limited assumptions on the model or its parameters. 

The proposed clustering technique synthesizes mobility data quantitatively (number of trips) and spatially (mean trip origin and destination). This allows identifying recursive mobility patterns that - when combined with weather data - provide accurate predictions of the demand. 

The method is tested with real-world data from New York City. We synthesize more than four million trips into vectors of movement, which are then combined with weather data to forecast the daily demand at a city-level. 
Results show that, already with a one-parameters model, the proposed approach provides accurate predictions.  
\footnote{This research has been partially sponsored by the European Union’s Horizon 2020 research and innovation programme under the Marie Skłodowska-Curie grant agreement No 754462 and the H2020 project NOESIS - No769980}
\end{abstract}

\rfoot{isbn}

\section{Introduction}

Bike Sharing Systems (BSS) proved to be an effective scheme to complement public transportation and car sharing services. If usually this mobility solution is associated to sustainable urban development, reduction in greenhouse gases, health benefits and reduction of on-road vehicles, recent studies show that this scheme also brings significant economic benefits for the urban economy \cite{c1}. Properly designed BBS can, in fact, improve spatial-connectivity of transport systems and deliver time-savings that far exceed commonly claimed benefits \cite{c1}.

Like for other transport modes, the success of BSS depends on an optimal balance between the supply and the demand. As spatial and temporal fluctuations of bike rentals lead to a sub-optimal distribution of bikes between different urban areas, the main challenge to handle BSS in an efficient way is to understand the underlying structure of its demand avoiding supply imbalance  \cite{c2,c3}.

This paper contributes to the existing research on this topic by introducing a new approach to forecast the expected BSS demand. First, we synthesize mobility data into vectors to identify similar daily mobility patterns and understand the systematic mobility pattern of a day-type. Then, we refine this general classification using contextual data (e.g. weather). Under this assumption, we develop a framework to infer recursive behaviour from a historical database of bike-sharing trips and to use this information to predict the daily demand for BSS.

While different approaches have already been proposed in the literature, the proposed methodology brings two main contributions. First, it is a low dimensional approach. This means that the number of parameters to be calibrated in order to achieve a good estimation is limited. In this paper, satisfactory results are achieved with a one-parameter model. Second, we show that – if weather data are included in the model – prediction capabilities can largely improve. 
We illustrate the method with publicly available trip data from the New York public bike system.  We collect and synthesize trip data from summer 2016 (over 4 million bike trips) and use these data to make an accurate prediction of the  daily demand for BSS service at a city level.

\section{Literature Review}
Bike Sharing Systems (BSS) can be grouped into two types: station-based and free-floating. In the first case, people can rent a bike from a certain dock – or station – and deliver it to a different dock belonging to the same operator. In the second case, users can leave their bike wherever they want, removing the need for a specific station/infrastructure \cite{c2,c4}.  \cite{c4}.

\vspace*{10px}
\small{\textbf{Article:} presented at}\\
\small{6th International Conference on Models and Technologies for Intelligent Transportation Systems}\\
\small{© 2019 IEEE.  Personal use of this material is permitted.  Permission from IEEE must be obtained for all other uses, in any current or future media, including reprinting/republishing this material for advertising or promotional purposes, creating new collective works, for resale or redistribution to servers or lists, or reuse of any copyrighted component of this work in other works}
\small{\textit{DOI:10.1109/MTITS.2019.8883283}}\\
\small{\textit{Copyright: 978-1-5386-9484-8/19/\$31.00 ©2019 IEEE}}

\normalsize	{}

This system has two main advantages: first, it drastically reduces start-up costs by removing the burden of building new docking stations. Second, it allows more flexibility for the user who can drop the bike close to his/her destination. However, this flexibility comes with the major drawback of making rebalancing operations even more complex and ill-predictable

The last decade has witnessed an intensive research effort to tackle these issues and provide optimal design and rebalancing strategies for BSS. From the strategical point of view, researcher mostly focused on the optimal network design problem, which includes optimizing number and location of dock stations in the BSS \cite{c5} or infrastructure design \cite{c6}.

A second branch of research focuses instead on the optimal BSS management. In order to keep a high level of service, BSS operators need to ensure a certain distribution of bikes among different docking stations. However, during the day, this distribution changes, leading to a lower level of service. A redistribution operation is thus required in order to re-establish optimality. This process is known in the literature as Bike Rebalancing problem \cite{c7}. Models dealing with this issue can be classified according to \textit{strategy} and \textit{type} \cite{c4}. Rebalancing strategies are mostly divided into operator-based or user-based strategies. In the first case, the operator leverages a fleet of vehicles to redistribute bikes among different stations while in the second case users are encouraged to self-balance the system \cite{c4}. Concerning the type, rebalancing strategies are divided into static and dynamic. In the former case the redistribution operation is performed when the system is not operating (for instance at night) while in the latter is performed in real-time \cite{c3}. 

Finally, many works focus on predicting demand patterns of bike sharing systems. Demand is, in fact, a fundamental input for the rebalancing problem. These models – called Bike Sharing Demand Prediction models – can be classified based on their spatial granularity according to three main groups: \textit{City-level}, \textit{Cluster-level} and \textit{Station-level} \cite{c2}

As the name suggests, the difference between these three groups depends on the precision level of the prediction model. In the first case – \textit{City-level} - the goal is to predict the overall demand for a large urban area \cite{c8}. \textit{Cluster-level} models assume that stations within a certain group (not necessarily within the same area) are correlated. Consequently, the demand prediction model estimates the demand for each cluster by assuming that the demand within the cluster will self-equilibrate– i.e. users will find at least one station with an available bike \cite{c9}. Finally, \textit{Station-level} approaches are the most precise, as they aim at predicting the demand for every single station in the system \cite{c10}. The main advantage is that correlations between different docks can be estimated and used to have more reliable and interpretable results \cite{c2}. On the other hand, given a certain number of observations, the potential estimation error increases with the number of variables \cite{c11}. Additionally, these solutions are not applicable to free-floating BSS. 

To conclude big datasets of bike data are often used in system control and optimization (e.g. rebalancing), but their utilization in demand analyses is still not fully exploited. In this paper we show how to synthesize big mobility data and how to utilzie it in sytem performance predictions.

\section{Methodology}

\subsection{The methodology at a glance}
In this work, we assume that the demand for bike-sharing services can be modelled as the combination of two components: a \textit{systematic} component, composed of highly predictable travel patterns identified at the clustering level, and a \textit{non-systematic} component, which is highly irregular and ill-predictable. Under this assumption, this section introduces a new framework to infer recursive behaviour from a historical database of bike-sharing trips. Contextual data (weather data) are then used to study their heterogeneity and daily variability and to predict the demand for a group of stations. 

\begin{figure}[!tb]
	\includegraphics[width=\columnwidth,keepaspectratio]{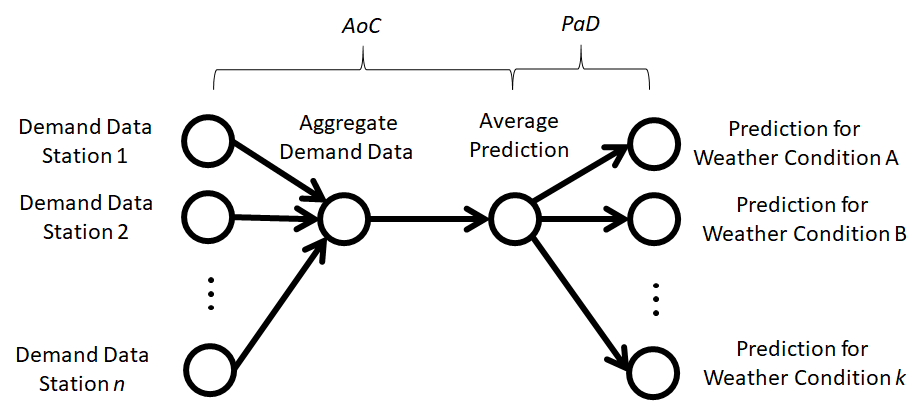}
	\caption{LS-BSS conceptual framework}
	\label{fig:fig1}
	\vspace*{2ex}
\end{figure}

The conceptual framework of the proposed Low Dimensional (LD) model for BSS  demand forecasting - simply called LD-BSS in the rest of this paper - is showed in Figure \ref{fig:fig1}. In essence, for a given group of stations and trip data, the so-called \textit{Vectors of movements} - formally defined in the next subsection - are calculated. Then, days with similar mobility patterns (indetified through similarity of above vectors) are grouped to provide generic prediction of the systematic demand based on the day-type (working, non-working, ...). Finally, additional information about historical weather condition is used to study the variability of the demand and provide weather-specific demand predictions. 

We can thus break down the proposed approach into two main steps, named \textit{Aggregation and Clustering} and \textit{Prediction and Disaggregation}.

\mytitle{Aggregation and Clustering (AaC)} In this phase, we synthesize mobility data into “so-called” vectors of movement.  They connect the centre of gravity of trip origins with the centre of gravity of trip destinations in AM and PM peaks respectively.  This synthetic representation of mobility allows reducing the set of recorded trips into a compact structure convenient for further processing.  Such vector formulation allows for comparison with classical similarity measures (e.g. cosine similarity). Thanks to this, we can formulate the clustering problem and build the similarity measure between mobility patterns.

\mytitle{Prediction and Disaggregation (PaD)} The \textit{AaC} phase groups historical observations on bike-sharing trips in clusters of similar day type. However, for the same day type, different demand values can be observed. This is an expected output, as the demand for bike sharing systems not only fluctuates within the day but also changes with respect to other phenomena, among the other, season and weather conditions \cite{c9}. As a consequence, the prediction model identifies the most likely “vector of movement” for a given season and meteorological condition, which are well-known to be the main elements influencing the bike sharing demand, and disaggregate this information in demand values that can be used to provide weather specific demand forecasting \cite{c9,c2}. 

\subsection{Data}

Generic record in the mobility pattern is a trip $T_i$ \eq{trip}. 
While the complete trip description is complex and might include a full path in space and time, the proposed method is suited for typically available minimal description of trip origin $O_i$ destination  $D_i$, start time $t_i$ and duration $\Delta T_i$. 
Such representation can be obtained both from classical travel survey and from automatic data collection systems (e.g. bike-sharing, car-sharing or taxi data) exploited in this paper.
Trips origins and destinations are understood as their spatial coordinates and are  often predefined, e.g. in station-based BSS where bicycles can be rented at stations only
while trip duration is given explicitly, the trip distance can only be inferred since the exact route is not given.
Due to privacy issues, the personal information are not recorded and personal trip pattern cannot be traced.

\begin{equation}\label{trip}
T_i=\lbrace O_i,D_i,t_i, \Delta t_i \rbrace 
\end{equation}

Trips recorded over a given time period (day, hour, or week) form a set of observations \eq{set}. In this paper we illustrate the proposed LD-BSS with daily mobility patterns, i.e. set of trips conducted during a given day:

\begin{equation}\label{set}
M_i = \lbrace T_1, T_2, \dots , T_n \rbrace
\end{equation}

\subsection{Synthesizing mobility pattern: Vectors of movements}

Numerous trips, comprising a mobility pattern form a complex system. Typical cardinalities are thousands or even millions of trips, with various origins, destinations and start times and durations. 
It is far from obvious when two mobility patterns are similar. 

We presume the fundamental set features like cardinality, total and mean trip duration are not enough to identify mobility pattern similarities and dissimilarities. Therefore we propose the following mapping.

We start from the concept of \textit{gravity centre} (mass centre), an arithmetic mean of trip origins and/or destinations.
Then we introduce \textit{vectors} spanning between them.
Due to the particular meaning of peak hours in the mobility patterns, we introduce a vector for \textit{AM} and \text{PM} peak hours.
For generic mobility pattern $M$ we introduce centre of gravity for origins \eq{center_O} and destinations \eq{center_D} and the vector of movement \eq{vector} spanned between them. 

\begin{equation}\label{center_O}
O_M = E(O_i : i \in M)
\end{equation}
\begin{equation}\label{center_D}
D_M = E(D_i : i \in M)
\end{equation}
\begin{equation}\label{vector}
\vec{V} = \overrightarrow{O_M D_M} 
\end{equation}

From the daily mobility we analyse trips of the $AM$ and $PM$ peaks (which are most important). Peaks are identified from the average recorded temporal profile as the two busiest morning and afternoon hours. From mobility pattern two subsets are selected: $M_{AM}=\lbrace T_i : T_i \in M, t_i \in AM \rbrace$, and $M_{PM}=\lbrace T_i : T_i \in M, t_i \in AM \rbrace$ respectively. Two vectors of movement computed for the two subsets of mobility pattern for the mapping utilized in the method:
\begin{equation}
M \rightarrow \lbrace \vec{V}_{AM} , \vec{V}_{PM}\rbrace
\end{equation}
In fact, the above mapping transforms any number of trips into four points: \textit{AM} origin and destination, \textit{PM} origin and destination. Such interpretation synthesizes all main characteristics of mobility patterns.

\subsection{Similarity measure}
With such representation, pairwise comparison of days, which was troublesome for a set of recorded trips, becomes possible. 
We propose to compare two generic vectors  $\vec{V}$ and $\vec{V'}$ with a cosine similarity \eq{cos} which returns similarity from range 0 to 1, 1 for vectors of equal length and direction and 0 for either orthogonal vectors or vectors of different lengths. Importantly, cosine similarity does not use the actual location. Nonetheless, it happens to be sensitive to the day type.

\begin{equation}\label{cos}
S(\vec{V},\vec{V'}) = \frac{\vec{V}\cdot \vec{V'}}{\vert \vec{V} \vert \vert \vec{V'} \vert}
\end{equation} 

\smallskip
Consequently, we can introduce the pairwise distance measure between two days \textit{N} and \textit{M}: 

\noindent
\begin{equation}\begin{aligned}\label{eq:similarity}
        \begin{split}
        \begin{array}{l@{\qquad}l}
            d(N,M) = \\[4pt]
            \alpha \cdot S(\vec{M}^{AM},\vec{N}^{AM}) + (1-\alpha) \cdot S(\vec{M}^{AM},\vec{N}^{AM}) \\[4pt]
            \end{array}
    \end{split}\hspace{-10pt}\end{aligned}
\end{equation}

\smallskip
With $\alpha$ being normalized weight, treated as a parameter of the procedure (we use default $\alpha=0,5$ in the case-study). If contextual data are directly combined with cosine similarity in order to obtain a pairwise distance that takes into account both day type and weather condition, \eq{eq:similarity} becomes:  

\smallskip

\begin{equation}\begin{aligned}\label{eq:full_similarity}
    \begin{split}\begin{array}{l@{\qquad}l}
    \resizebox{.18\hsize}{!}{$d(N,M) =$} \\[4pt]
    \resizebox{.79\hsize}{!}{$\beta_1\cdot[\alpha \cdot S(\vec{M}^{AM},\vec{N}^{AM}) + (1-\alpha) \cdot S(\vec{M}^{AM},\vec{N}^{AM})]$} \\[4pt]
    \resizebox{.29\hsize}{!}{$+ \beta_2\cdot S(\vec{\theta}^{M},\vec{\theta}^{N})$} \\[4pt]
    \end{array}
    \end{split}\end{aligned}
\end{equation}

\smallskip

Where $\beta_1$ and $\beta_2$ are weights assigned to each component depending on the trust that one has on the data while  $S(\theta^{M},\theta^{N})$ represents the similarity between weather conditions during the two days. In this paper, we use the average daily temperature as a proxy for weather conditions and the similarity is calculated as:

\begin{equation}\label{eq:Sim_weather}
S(\theta^{M},\theta^{N}) = \frac{|\theta^{M}-\theta^{N}|}{\theta^{N}}
\end{equation} 

With $\theta^M$ and $\theta^N$ the average temperature for day $M$ and $N$, respectively. Such metrics can be applied for most of the clustering methods. 

\subsection{Clustering}
Clustering procedure yields cluster membership map, i.e. labels each day with a cluster id $c(M_i)$, consequently the cluster is the subset of days belonging to a given cluster $C=\lbrace M_i : c(M_i) = C \rbrace $.

In this research, we apply the \textit{agglomerative hierarchical clustering algorithm}. Specifically, we exploit the implementation proposed in \textit{Scikit-learn}, which is an open-source library developed in Python \cite{c12}.
Yet any alternative method can be applied.

\subsection{Cluster decomposition}\label{sec:dec}
If a sufficient number of observations is available, results from the clustering procedure can be adopted to forecast the mobility demand $M_c$ for current cluster and associated day-types (working day, holiday, etc.) with  \eq{pred}.

\begin{equation}\label{pred}
M_p = E(M_i : i \in C)
\end{equation}

However, the predicted mobility patter $M_c$ for a given day should change significantly according to weather conditions and seasonality. This phenomenon is not explicitly modelled if \eq{eq:full_similarity} is used within the clustering procedure. We thus include additional information concerning the average temperature to further classify mobility patterns based on the average temperature of each day. 

In this work, we adopt the concept of temperature class $\Theta_c$ to combine weather data within the proposed LD-BSS framework. Let us define $\theta_{i}$ as the average temperature related to a certain set of observations $M_i$. Let us also define $\theta_{low}$ and $\theta_{up}$ as the minimum and maximum average temperature for a certain class $\Theta_c$, respectively. Then, each observation $M_i$ can be associated to a certain class, as showed in \eq{eq:temp_class}: \smallskip\indent
\begin{equation}
C^{\Theta_c} = \lbrace \theta_{low} < \theta_{i} \leq \theta_{up} : i \in \Theta_c)
\label{eq:temp_class}
\end{equation}

Where $C^{\Theta_c}$ represents the subset of observations belonging to the temperature class $\Theta_c$. The improved prediction model can then be written as:

\begin{equation}
C^* = C \cap C^{\Theta_c}
\end{equation}

\begin{equation}
M_p = E(M_i : i \in C^*)
\end{equation}

Which provides the most likely mobility pattern for a given cluster and temperature. 
\subsection{Validation}
Obtained results may be verified at the two levels. First, by internally looking at the clustering quality, i.e. the within-cluster consistency and between-cluster separation. This may be observed by looking at the clustering quality measures, like silhouette score \cite{c13}. Apart from this the proposed method is validated by assessing quality of its predictions, namely by looking at how well the actual (empirically observed) system variables are reproduced on the test set using predictions made from the train set. The performance measure adopted in this study is the Root Mean Squared Error (\textit{RMSE}):

\begin{equation}
    RMSE_i=\sqrt{\frac{1}{N}\sum\limits_{t=1}^N{(\hat{y}_i^t-y_i^t)^2}}
\end{equation}

Where $\hat{y}_i^t$ and $y_i^t$ are respectively the predicted and observed number of trips for day $i$ and time interval $t$. 

\section{Case Study}

We test the proposed LD-BSS method with publicly available trip data from the New York City bike system. We collect and synthesize trip data between 01/06/2016 to 30/09/2016 (over 4 millions bike trips). City Bike NYC allows using one of over 12 000 bicycles to travel between more than 750 stations located in New York City and Jersey City, New Jersey. Basic statistics are presented in Table \ref{tab:1}.

\begin {table}[H]
\caption {Dataset used to test and illustrate the method.} \label{tab:1} 
\begin{center}
\begin{tabular}{l|l} 
\toprule
	{first} & {2016-06-01}  \\
    {last} & {2016-06-30}  \\
    {\#stations} & {831}  \\
    {\#bicycles} & {14 851}  \\
    {\#trips} & {4 034 347}  \\
    {\#daily trips avg} & {47 462}  \\
    {-- min} & {22 397}  \\
    {-- max} & {58 674}  \\
     \bottomrule
\end{tabular}
\end{center}
\end {table}

\begin{figure*}[!tb]
	\centering
	\includegraphics[scale=1]{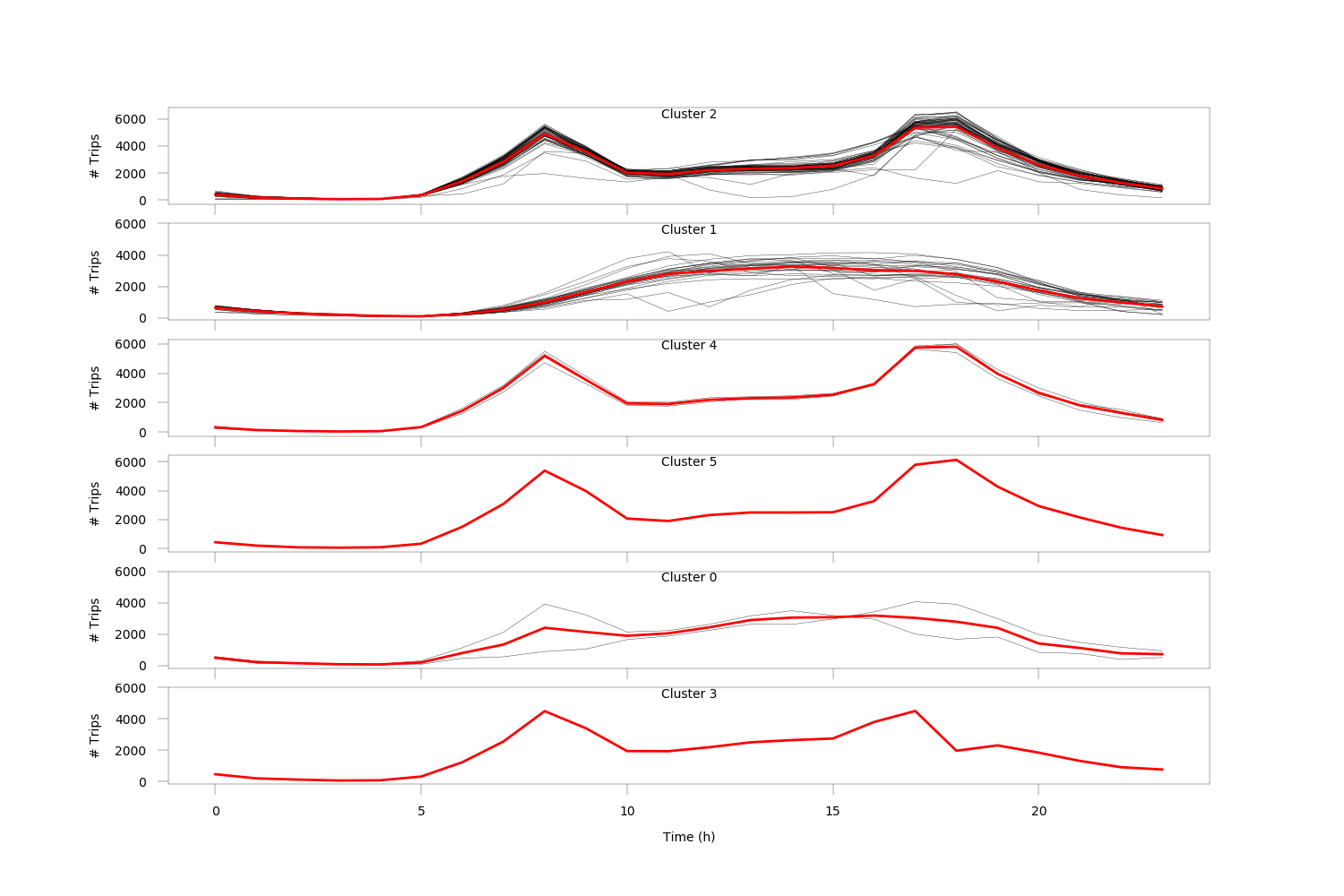}%
	\caption{Demand Profiles within each Cluster - Observations (Black) and Prediction (Red)}%
	\label{fig:cluster}%
	\vspace*{-1ex}%
\end{figure*}

\begin{figure*}[!tb]
	\centering
	\includegraphics[scale=1.1]{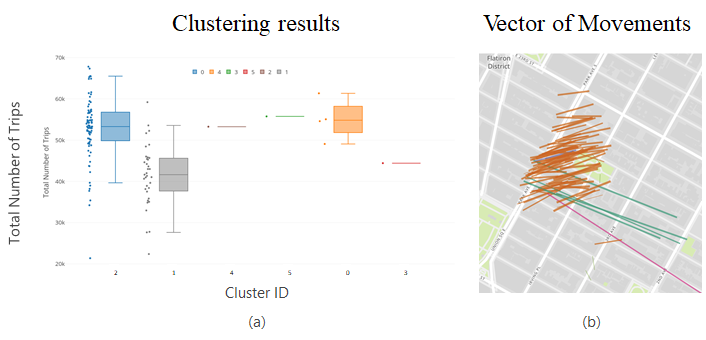}%
	\caption{(a) Within-Cluster distribution of the demand; (b) Vector of Movements}%
	\label{fig:fig3}%
	\vspace*{-1ex}%
\end{figure*}

For each trip information about its origin, destination start and end time is available.

For the same time period, historical weather data are also available\footnote{Source: https://w2.weather.gov/climate}. In this study, data about the average temperature in New York Central Park area have been assumed to be representative of the entire study area. While more detailed information is available (higher spatial accuracy, precipitation data, snow depth), this study is an exploratory analysis aimed at answering two questions. First, if the proposed low dimensional model can provide accurate predictions of the demand for a given database. Second, the influence of weather data within the proposed framework. For these reasons, we focused on investigating the relationship between temperature and trip data, leaving spatial granularity for later studies. 

In order to answer these two questions, the database has been divided into two parts. About 70\% of the data - 85 days in total - have been used to train the model and to generate the prediction model.  Then, this model has been adopted to predict the demand for the reaming 37 days. Two different experiments have been performed to investigate the relationship between vectors of movements and weather data. First, to use only vector of movements during the training phase and to leverage the decomposition scheme proposed in Section \ref{sec:dec} to further improve predictions. Then, we investigate the effect of combining weather data and vectors of movements within the clustering, as showed in \eq{eq:full_similarity}. On the one hand, this solution simplifies the model by removing the decomposition phase. On the other hand, additional parameters need to be properly calibrated in order to provide accurate predictions.            

\subsection{Aggregation and clustering}

This section introduces the results from the clustering procedure, meaning clustering map and average prediction model.

The aggregation and clustering phase estimated six typical day-types. Figures \ref{fig:cluster} and \ref{fig:fig3}a show the clustering results, while Figure \ref{fig:fig3}b graphically illustrates how these vectors of movements look like.

Concerning the clustering, results clearly identify two main groups of observations. Cluster (2) contains observations about typical behaviour during working days (Monday to Friday), while Cluster (1) mostly represents weekends and public holidays. Most importantly, Clusters (1-2) are consistent in terms of parameters not-included in the clustering procedure but significant to mobility, such as the total number of trips and temporal profile (as visible on \ref{fig:cluster} and \ref{fig:fig3}a). Clusters (0,3,4,5) represent outliers, typical for agglomerative clustering technique used in the paper. They cannot be used for prediction purposes, as this approach captures only systematic behaviour. 

While Figure \ref{fig:cluster} supports the claim that vectors of movements are sufficient to infer human mobility and detect mobility patterns, Figure \ref{fig:fig3} shows that demand profiles within a cluster largely vary. This means that the average prediction model - marked with the red line in Figure \ref{fig:cluster} - could provide a biased prediction of the daily demand patterns. Additional information is required to reduce the variance within each cluster and provide accurate predictions.   

\subsection{Disaggregation}

In this section, we show how the decomposition scheme proposed in Section \ref{sec:dec} reduces results variability without introducing additional parameters. 
In order to generate the refined cluster $C^*$, temperature data have been divided into six classes $\Theta$: 
\begin{equation}
    \begin{split}
        \begin{array}{c@{\qquad}c}
        0.000^\circ F < \Theta_1 \leq 57.50^\circ F\\[4pt]
        57.40^\circ F < \Theta_2 \leq 65.25^\circ F\\[4pt]
        65.25^\circ F < \Theta_3 \leq 73.00^\circ F\\[4pt]
        73.00^\circ F< \Theta_4  \leq 80.75^\circ F\\[4pt]
        80.75^\circ F < \Theta_5 \leq 88.50^\circ F\\[4pt]
        88.50^\circ F < \Theta_6 
        \end{array}
    \end{split}
\end{equation}

\begin{figure}[!tb]
	\includegraphics[width=\columnwidth,keepaspectratio]{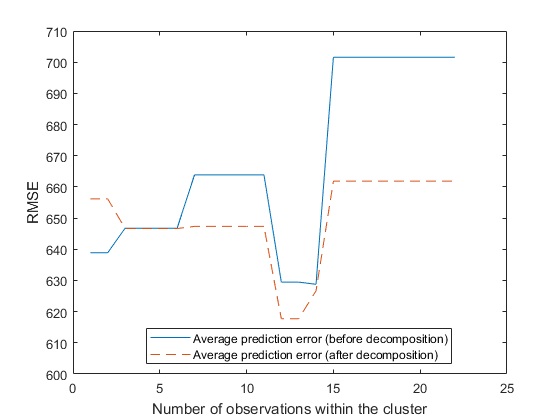}
	\caption{Average prediction error as a function of the number of observations}
	\label{fig:decomp}
	\vspace*{2ex}
\end{figure}

By accordingly decomposing each cluster, a total of 22 refined clusters $C^*$ have been obtained. Figure \ref{fig:decomp} shows the results and, specifically, the average prediction error in terms of Root Mean Square Error (\textit{RMSE}). The $x$ axis represents instead the precision of the model. To each refined cluster $C^*$ corresponds, in fact, a specific prediction model. However, when few observations for a certain temperature class $\Theta_c$ or for a certain day type $C$ are available- the decomposition process can return a refined cluster $C^*$ with one or zero elements. This means that the prediction model will also be calculated only on a very limited amount of data.

In Figure \ref{fig:decomp}, \textit{Number of observations within the cluster} equals to one means that even refined clusters $C^*$ with one single observation have been used to forecast the demand for BSS. By contrast, a \textit{Number of observations within the cluster} equals to 5 means that only refined clusters $C^*$ with more than 5 observations have been accepted within the improved prediction model, while the others have been discarded.  

We can observe that, when the average error over all days is computed (\textit{Number of observations within the cluster is 1}), the average RMSE is lower when weather data are not considered within the estimation process. However, this is reasonable, as this means that the historical database has limited or no information for that combination of day type and temperature. Thus, when weather data are not available, the normal average prediction based on all available data is providing a better estimation.

On the other hand, for clusters with more than 5 observations, the improvement becomes systematic. This suggests that, if enough weather data are available, the decomposition step can largely improve the prediction. In this specific case study, results show that the reduction is particularly effective for larger errors. When the average error is calculated for refined clusters with more than 15 elements, the largest improvement is achieved.  

\subsection{Sensitivity analysis}

\begin{figure}[!tb]
	\includegraphics[width=\columnwidth,keepaspectratio]{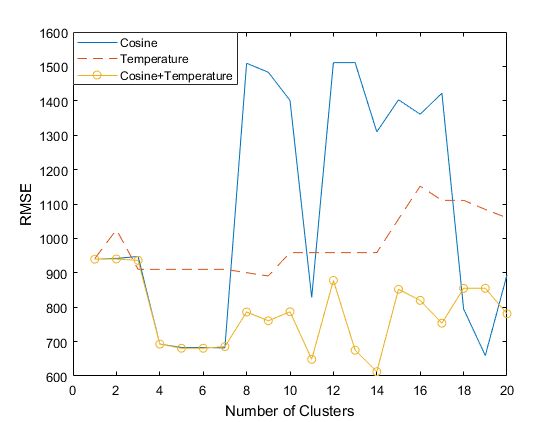}
	\caption{Average prediction error as a function of the parameters}
	\label{fig:sensitivity}
	\vspace*{2ex}
\end{figure}

The model discussed in the previous sub-sections is a one-parameter model. The only parameter is in fact the number of clusters, which was assumed equal to 6. In this section, we evaluate model performances for different number of clusters. Additionally, we include weather information within the clustering phase to investigate the difference between using the decomposition step and relying on the clustering to directly infer refined clusters $C^*$. We thus study the average prediction error (RMSE) in three cases: (i) only vector of movements are used to create the average prediction model [\textit{Cosine similarity}] \eq{eq:similarity}, (ii) only weather data are used [\textit{Temperature}] (\eq{eq:full_similarity} and $\beta_1=0$) and, finally, (iii) their combined effect [\textit{Cosine+Temperature}] (\eq{eq:full_similarity} and $\beta_1,\beta_2\neq0$).  Results are shown in Figure \ref{fig:sensitivity}.    

Concerning the base-case, the first conclusion is that - for the database adopted in this study - a number of clusters between four and seven is convenient when \textit{cosine similarity} is the sole similarity measure. Since vectors of movements mostly identity day-type and systematic mobility patterns, when too many clusters are created, the model over-fits available observations losing its predicting capabilities. 

Weather data, on the other hand, provide a relatively poor estimation. For a small number of clusters, the error is significantly higher than for both other approaches and it slowly increases for an increasing number of clusters. This suggests that weather data alone are not suited to predicting BSS demand. 

Finally, when weather data and vectors of movements are combined together, the model provides the best performance. \textit{Cosine} and \textit{Cosine+Temperature} show almost the same performance for a number of clusters equal or less than seven. The reason is that a higher number of clusters is needed to leverage weather data. In the previous experiment, starting from an initial set of six clusters, 22 refined clusters have been identified during the decomposition phase. As a consequence, when the number of clusters increases, the combined approach takes advantage of the additional information to avoid overfitting the data. 

An additional consideration can be done by comparing Figures \ref{fig:decomp} and \ref{fig:sensitivity}. When weather data are used within the clustering algorithm, the proposed LD-BSS provides overall a better estimation. The average error is in fact calculated as the average error over all 37 days and it is still lower than the best result showed in Figure \ref{fig:decomp}. However, when weather data and vectors of movement are combined together, weights $\beta_1$ and $\beta_2$ need also to be calibrated. If these parameters are not calibrated together with the number of clusters, prediction capabilities of the model decrease drastically. On the other hand, the decomposition scheme allows a systematic improvement of the results without introducing additional parameters.

\section{CONCLUSIONS}

This paper introduces a Low Dimensional (\textit{LD}) model for Bike Sharing demand forecasting. The proposed model leverages a two phases approach to exploit available information while keeping the number of parameters to be tuned low. 

There are two main innovative elements First, we propose a new method to synthesize big, multidimensional trip data sets. This synthetic description of daily mobility through vectors of movement (morning and afternoon) significantly reduces the problem size and allows to introduce pairwise distance measure and, consequently, to apply clustering methods, which was not possible on the raw trip data. Second, we combined said vectors of movements with temperature data - which is assumed to be a proxy for weather conditions in this study. We show that the combined effect of these two elements can substantially improve the accuracy of the prediction model. 

Two ways of combining vectors of movements and weather data have been investigated. First, we propose a two-phase approach, which allows to include additional data without increasing the number of parameters of the model. Then we investigate the possibility of combining weather data within the clustering algorithm. Both approaches provide satisfactory estimations. 

This paper applied the proposed methodology with publicly available trip data from the New York City bike system to forecast BSS demand at a city-level. However, the same approach can be adopted to clusters, thus guaranteeing a higher spatial granularity.  

Future work will focus on investigating and combining more weather data (precipitation data, snow depth, ...) and a higher spatial granularity.

\addtolength{\textheight}{-12cm}   




\end{document}